\documentclass[aps,prb,superscriptaddress,twocolumn,showpacs]{revtex4}

\usepackage{graphicx}
\usepackage{dcolumn}
\usepackage{bm}
\usepackage{amsmath}

\begin{document}

\title{Low-temperature heat transport of CuFe$_{1-x}$Ga$_x$O$_2$ ($x =$ 0--0.12) single crystals}

\author{J. D. Song}
\affiliation{Hefei National Laboratory for Physical Sciences at Microscale, University of Science and Technology of China, Hefei, Anhui 230026, People's Republic of China}

\author{X. M. Wang}
\affiliation{Hefei National Laboratory for Physical Sciences at Microscale, University of Science and Technology of China, Hefei, Anhui 230026, People's Republic of China}

\author{Z. Y. Zhao}
\affiliation{Hefei National Laboratory for Physical Sciences at Microscale, University of Science and Technology of China, Hefei, Anhui 230026, People's Republic of China}
\affiliation{Fujian Institute of Research on the Structure of Matter, Chinese Academy of Sciences, Fuzhou, Fujian 350002, People's Republic of China}

\author{J. C. Wu}
\affiliation{Hefei National Laboratory for Physical Sciences at Microscale, University of Science and Technology of China, Hefei, Anhui 230026, People's Republic of China}

\author{J. Y. Zhao}
\affiliation{Hefei National Laboratory for Physical Sciences at Microscale, University of Science and Technology of China, Hefei, Anhui 230026, People's Republic of China}

\author{X. G. Liu}
\affiliation{Hefei National Laboratory for Physical Sciences at Microscale, University of Science and Technology of China, Hefei, Anhui 230026, People's Republic of China}

\author{X. Zhao}
\affiliation{School of Physical Sciences, University of Science and Technology of China, Hefei, Anhui 230026, People's Republic of China}

\author{X. F. Sun}
\email{xfsun@ustc.edu.cn}

\affiliation{Hefei National Laboratory for Physical Sciences at Microscale, University of Science and Technology of China, Hefei, Anhui 230026, People's Republic of China}

\affiliation{Key Laboratory of Strongly-Coupled Quantum Matter Physics, Chinese Academy of Sciences, Hefei, Anhui 230026, People's Republic of China}

\affiliation{Collaborative Innovation Center of Advanced Microstructures, Nanjing, Jiangsu 210093, People's Republic of China}

\date{\today}

\begin{abstract}

We report a study on the thermal conductivity of CuFe$_{1-x}$Ga$_x$O$_2$ ($x =$ 0--0.12) single crystals at temperatures down to 0.3 K and in magnetic fields up to 14 T. CuFeO$_2$ is a well-known geometrically frustrated triangular lattice antiferromagnet and can be made to display multiferroicity either by applying magnetic field along the $c$ axis or by doping nonmagnetic impurities, accompanied with rich behaviors of magnetic phase transitions. The main experimental findings of this work are: (i) the thermal conductivities ($\kappa_a$ and $\kappa_c$) show drastic anomalies at temperature- or field-induced magnetic transitions; (ii) the low-$T$ $\kappa(H)$ isotherms exhibit irreversibility in a broad region of magnetic fields; (iii) there are phonon scattering effect caused by magnetic fluctuations at very low temperatures. These results demonstrate strong spin-phonon coupling in this material and reveal the non-negligible magnetic fluctuations in the ``ground state" of pure and Ga-doped samples.

\end{abstract}


\maketitle
\section{INTRODUCTION}

Multiferroicity, in which magnetism and ferroelectricity co-exist, is a result of strong coupling between magnetic and electric degrees of freedom in insulators and has attracted much attention because of its potential applications, since the electric (magnetic) properties of this kind of material can be modulated by an external magnetic (electric) field.\cite{Schmid, Kimura1, Hur, Lottermoser, MF_Review} The delafossite CuFeO$_2$ is one of the candidates for such magnetic ferroelectrics.\cite{Kimura2, Arima}

In CuFeO$_2$, the Fe$^{3+}$ ($S =$ 5/2, $L =$ 0) ions are the only magnetic elements and they antiferromagnetically interact with each other, forming a good example of triangular lattice antiferromagnets (TLAs). Since the orbital singlet Fe$^{3+}$ ions should have Heisenberg character, a non-collinear magnetic ground state with three spins aligned at 120$^\circ$ from each other in the base plane was naively expected.\cite{Muir, Kawamura} However, it was found that the low-temperature phase of CuFeO$_2$
is likely a collinear four-sublattice (4SL) state, adopting an in-plane up-up-down-down order with spins pointing along or anti-parallel to the $c$ axis.\cite{Mitsuda1, Mitsuda2, Ye1} In zero magnetic field, the crystal structure of CuFeO$_2$ undergoes lattice distortion from the hexagonal $R\bar{3}$m space group at $T_{N1} \sim$ 14 K, and gradually changes into a monoclinic $C2/m$ space group at $T_{N2} \sim$ 11 K.\cite{Terada1, Ye2, Terada2} In the meantime, the magnetic phase changes sequentially from the paramagnetic (PM) phase to the partially disordered incommensurate (PD or ICM) phase at $T_{N1}$ with a sinusoidally amplitude-modulated magnetic structure and the moment along the $c$ axis, and then undergoes a first-order transition at $T_{N2}$ to the 4SL state. Moreover, when a magnetic field is applied along the $c$ axis at $T < T_{N2}$, CuFeO$_2$ displays multi-step magnetic phase transitions.\cite{Kimura2, Petrenko, Terada3, Lummen} In an external magnetic field ($<$ 14 T), two first-order transitions occur and the field-induced spin structures have been described as follow:\cite{Kimura2} a collinear 4SL state for 0 $< \mu_0H <$ 7 T; a non-collinear incommensurate structure for 7 $< \mu_0H <$ 13 T with a proper helical magnetic order, in which ferroelectricity has been revealed, so this phase is called the ferroelectric incommensurate (FEIC) phase or multiferroic phase; and a collinear commensurate five-sublattice (5SL) state for $\mu_0H >$ 13 T, adopting an in-plane three-up two-down order with spin moments parallel (or anti-parallel) to the $c$ axis. The $H-T$ phase diagram can be summarized in Fig. 1(a).

Besides the magnetic-field-induced ferroelectricity, CuFeO$_2$ can also achieve a spontaneous ferroelectric phase in zero field by nonmagnetic Al$^{3+}$ or Ga$^{3+}$ doping on Fe$^{3+}$ site.\cite{Seki1, Terada4} Nonmagnetic doping causes lattice distortions and modification of magnetic interactions, which can realize the adjustment of magnetic ground state of CuFeO$_2$ by varying impurity concentration. For example, for Ga-doped CuFe$_{1-x}$Ga$_x$O$_2$, the magnetic ground state can transform into the FEIC phase with 0.018 $\le x \le$ 0.058.\cite{Terada5} In detail, in the case of $x =$ 0.035, with lowering temperature, the magnetic phase changes from the PM phase to an oblique partially disordered (OPD or ICM2) phase at $T_{N1} \sim$ 14 K, in which the magnetic moments are sinusoidally modulated along the $[110]$ axis and oriented by $\sim 50^{\circ}$ from the $c$ axis.\cite{Seki2, Terada6} With further lowering the temperature, the magnetic phase undergoes a second transition to the PD (or ICM1) phase at $\sim 11.5$ K.\cite{Seki2} Below $T_{N2} \sim$ 8 K, the magnetic phase is long-range ordered into the FEIC state. In this phase, a $c$-axis magnetic field can induce a magnetic transition to the collinear 5SL phase at $\mu_0H_c \sim$ 11 T.\cite{Seki2} With higher Ga-doping of 0.058 $\le x \le$ 0.08, the long-range magnetic order can be destroyed; instead, the OPD phase is established at low temperatures.\cite{Terada5, Nakajima1} The phase diagram of CuFe$_{1-x}$Ga$_x$O$_2$ is summarized in Figs. 1(b) and 1(c).

\begin{figure}
\centering\includegraphics[clip,width=6.5cm]{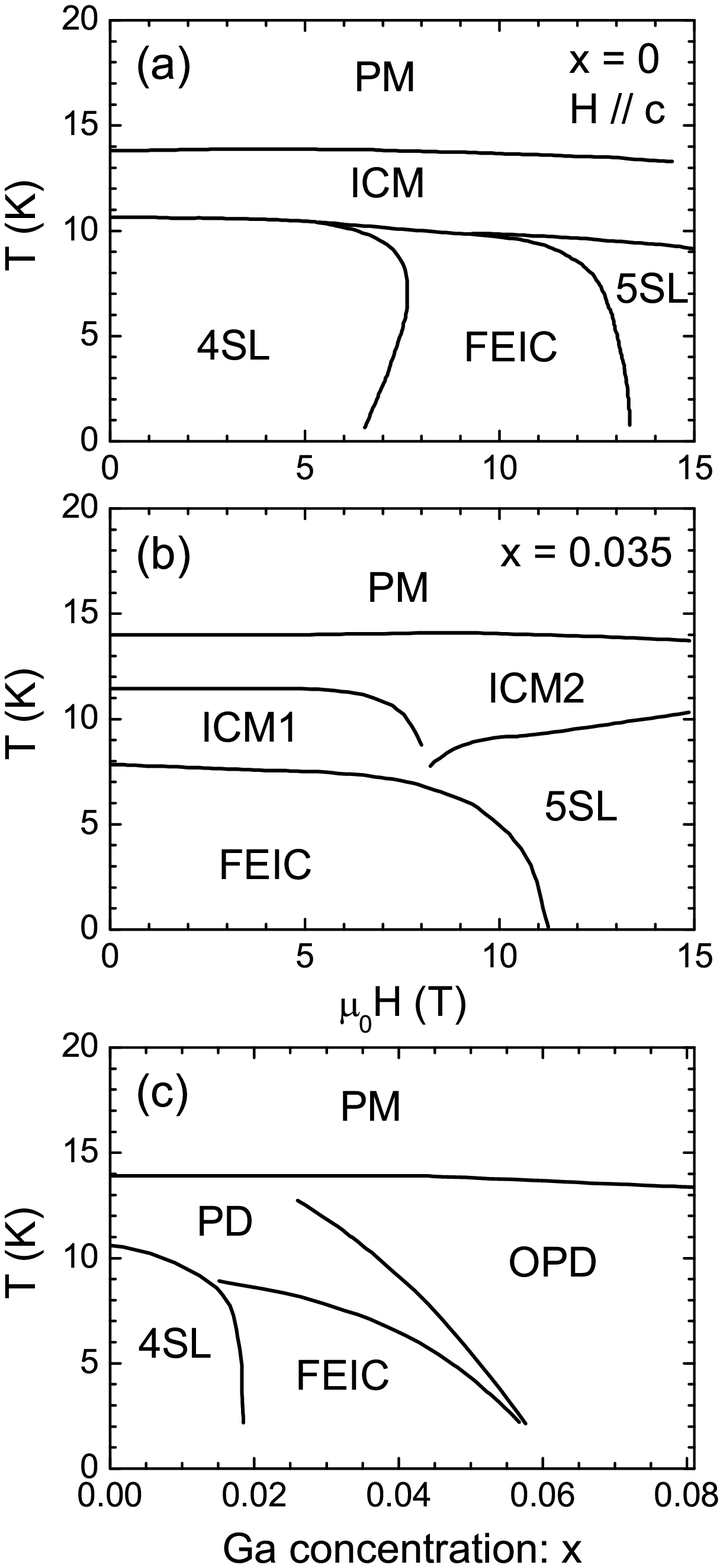}
\caption{(a,b) The $H-T$ phase diagrams of CuFe$_{1-x}$Ga$_x$O$_2$ with $x =$ 0 and 0.035 in $H \parallel c$. (c) The zero-field $T-x$ phase diagram of CuFe$_{1-x}$Ga$_x$O$_2$ with $x =$ 0--0.08. (Taken from Refs. \onlinecite{Kimura2, Seki2, Terada5}.)}
\end{figure}

As a basic physical property of solids, low-temperature heat transport has recently attracted much attention in the study of spin systems.\cite{Hess, Sologubenko, Zhao_SG, Yamashita, Sun_DTN, Zhao_IPA, Jeon} First, it is a useful probe of many kinds of elementary excitations such as phonon, magnon, and spinon.\cite{Berman, Ziman, Ashcroft} In the long-range-ordered magnetic insulators, magnons can act as heat carriers or phonon scatterers and thus affect the thermal conductivity.\cite{Sun_DTN, Zhao_IPA, Jeon} Second, the thermal conductivity is sensitive to the spin-phonon coupling, which is usually strong in multifferroic materials.\cite{Wang_HMO, Wang_TMO, Zhao_GFO, Zhao_DFO} As a result, low-temperature heat transport is an effective method to probe magnetic field induced transitions. Although CuFeO$_2$ has been studied for a long time, its heat transport properties have not been investigated. In this paper, we report a systematic study of the thermal conductivity ($\kappa$) of CuFe$_{1-x}$Ga$_x$O$_2$ ($x =$ 0--0.12) single crystals at low temperatures down to 0.3 K and in magnetic fields up to 14 T. Various magnetic transitions are detected by either the temperature-dependence $\kappa(T)$ or the field-dependence $\kappa(H)$ data, which demonstrate strong spin-phonon coupling in this system. In addition, based on unexplored sub-Kelvin-temperature thermodynamic and transport measurements, strong magnetic fluctuations in the ``ground state" were revealed.

\section{EXPERIMENTS}

High-quality CuFe$_{1-x}$Ga$_x$O$_2$ ($x =$ 0, 0.035, 0.08, and 0.12) single crystals were grown by using a floating-zone technique.\cite{Song_Growth} The chemical compositions were carefully determined by using X-ray fluorescence spectrometry (XRF) and inductively coupled plasma atomic-emission spectroscopy (ICP-AES).\cite{Song_Growth} Using X-ray Laue photographs, the large pieces of crystals were cut into long-bar shaped samples with specific orientations. The thermal conductivity was measured at low temperatures down to 0.3 K and in magnetic fields up to 14 T by using a conventional steady-state technique.\cite{Sun_DTN, Zhao_IPA, Wang_HMO, Wang_TMO, Zhao_GFO, Zhao_DFO} The heat current was along either the $a$ axis ($\kappa_a$) or the $c$ axis ($\kappa_c$), while the magnetic fields were applied along or perpendicular to the $c$ axis. The specific heat was measured by the relaxation method using a commercial physical property measurement system (PPMS, Quantum Design) equipped with a $^3$He insert.

\section{RESULTS AND DISCUSSION}

\subsection{Specific heat}

\begin{figure}
\centering\includegraphics[clip,width=6.5cm]{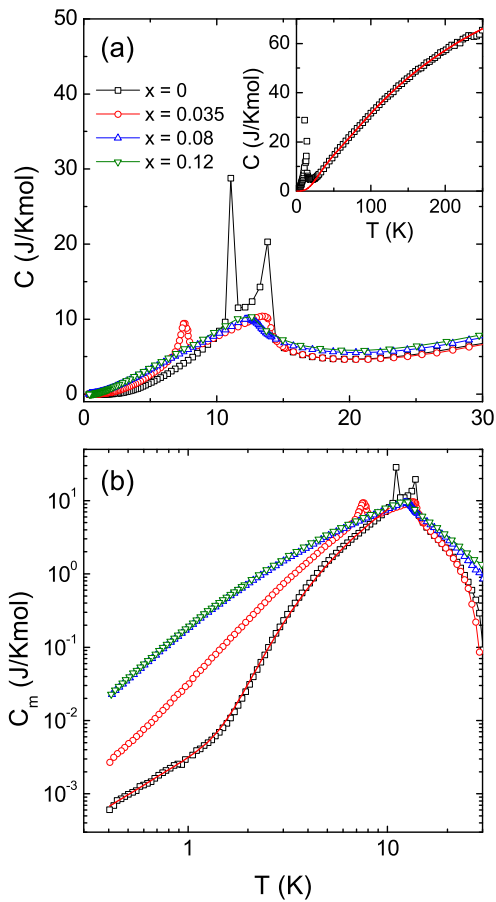}
\caption{(Color online) (a) Temperature dependence of the low-temperature specific heat of CuFe$_{1-x}$Ga$_x$O$_2$ ($x =$ 0, 0.035, 0.08, and 0.12) single crystals in zero field. Inset: the specific-heat data in a broader temperature range for the $x =$ 0 sample. The solid line shows the fitting result by using the modified formula (2) of the lattice specific heat, with the consideration of optical phonons. (b) The low-$T$ magnetic specific heat obtained by subtracting the calculated phonon contribution from the raw data. The solid line shows the fitting to the data by using formula (3).}
\end{figure}

The literature contains quite limited specific-heat data of pure and Ga-doped CuFeO$_2$.\cite{Petrenko, Terada4} The earlier studies were carried out for $x =$ 0 and 0.037 at temperatures down to 2 K. In this work, we measured the specific heat of CuFe$_{1-x}$Ga$_x$O$_2$ ($x =$ 0, 0.035, 0.08, and 0.12) single crystals at temperatures down to 0.4 K, as shown in Fig. 2, and revealed some new information. For the $x =$ 0 sample, the data exhibit two sharp peaks at $T_{N1} =$ 14 K and $T_{N2} =$ 11 K, which correspond to the second-order transition from the PM to PD (ICM) phase and the first-order transition from the PD (ICM) to 4SL phase, respectively.\cite{Kimura2, Terada1, Ye2, Terada2} For the $x =$ 0.035 sample, the lower-$T$ peak moves to 7.5 K ($T_{N2}$) and becomes weaker. The first transition is still located at 14 K ($T_{N1}$) but the peak is weak and very broad. These behaviors are consistent with earlier results.\cite{Petrenko, Terada4} In the case of $x =$ 0.08 and 0.12, there is only a broad peak at $T_N =$ 12.5 K, which should be related to the magnetic transition from the PM to OPD phase.\cite{Terada5} It is known that the specific-heat data can provide useful information about magnetic excitations. To analyze the data, we need to determine the phononic specific heat of CuFeO$_2$. For this purpose, the specific heat of the $x =$ 0 sample was measured at temperatures up to 250 K, as shown in the inset to Fig. 2(a); such measurement has not been reported in the earlier works. Initially, a fitting to the high-$T$ (above $T_{N2}$) data was tried by using the standard Debye formula,\cite{Tari}
\begin{equation}\label{eq:eps}
C_{ph} = 3 R N \left(\frac{T}{\Theta_D}\right)^3 \int_0^{\Theta_D /T} \frac{x^4e^x}{(e^x-1)^2}\mathrm{d}x,
\end{equation}
where $x = \hbar\omega/k_BT$, $R$ is the universal gas constant, and $N$ is the total number of acoustic phonon branches. In a simplified case, $N =$ 12 (considering 4 atoms in each unit formula). However, we found that the simple Debye formula cannot simulate the data accurately. One known reason for the deviations of high-$T$ specific heat from the Debye model is the contribution of optical phonons at high temperature, which can be described by the Einstein model.\cite{Svoboda, Hemberger, Janiceka} Actually, the phonon spectrum of CuFeO$_2$ should consist of 3 acoustic branches and 9 optical branches and inspired by this we found that the high-$T$ data can be fitted by the formula
\begin{equation}\label{eq:eps}
\begin{split}
C_{ph} = & \ 3N_D R \left(\frac{T}{\Theta_D}\right)^3 \int_0^{\Theta_D /T} \frac{x^4e^x}{(e^x-1)^2}\mathrm{d}x \\
& + N_{E1} R \left(\Theta_{E1}/T\right)^2 \frac{\mathrm{exp}(\Theta_{E1}/T)}{[\mathrm{exp}(\Theta_{E1}/T)-1]^2} \\
& + N_{E2} R \left(\Theta_{E2}/T\right)^2 \frac{\mathrm{exp}(\Theta_{E2}/T)}{[\mathrm{exp}(\Theta_{E2}/T)-1]^2}.
\end{split}
\end{equation}
Here, the first term is the contribution of 3 acoustic phonon branches using the Debye model ($N_D =$ 3), while the second and third terms are the contributions from the optical branches using the Einstein model ($N_{E1} =$ 5 and $N_{E2} =$ 4). The other parameters are the Debye temperature, $\Theta_D =$ 180 K, and two Einstein temperatures, $\Theta_{E1} =$ 440 K and $\Theta_{E2} =$ 1025 K. Note that the Debye temperature corresponds to a mean sound velocity of 2150 m/s, which is close to the experimental value determined by ultrasonic measurements.\cite{Quirion1}

The fitting results are taken as the lattice specific heat of CuFe$_{1-x}$Ga$_x$O$_2$. We can get the low-$T$ magnetic specific heat by subtracting the phononic term from the raw data, as shown in Fig. 2(b). Surprisingly, the $C_m(T)$ curve of the $x =$ 0 sample displays a distinct anomaly at 1.5 K. It should be pointed out that the magnetic properties of CuFeO$_2$ at such low temperatures have not been explored in the earlier works. Nevertheless, it is not likely a magnetic transition happening at 1.5 K since only the slope of $C_m(T)$ changes. The sudden change of the temperature dependence of specific heat can be directly related to a change of magnetic excitations. Actually, the data below $T_{N2}$ of this sample are well fitted by
\begin{equation}\label{eq:eps}
C_{m} = aT^n + bT\mathrm{exp}(-\Delta/T),
\end{equation}
with parameters $a =$ 3.08 $\times$ 10$^{-3}$ J/Kmol, $b =$ 2.13 J/K$^2$mol. $n =$ 1.67, and $\Delta =$ 10.6 K. This means that there are two kinds of magnetic excitations. The exponential term is due to gapped spectrum of the Ising-like 4SL phase.\cite{Terada7, Terada8, Ye1, Nakajima2, Nakajima3} In this regard, the neutron scattering has detected an anisotropy gap of 0.9 meV, which is very close to the 10.6 K gap from the present specific heat data.\cite{Ye1} This contribution to the specific heat quickly decays at $T < T_{N2}$. The power-law behavior dominates at very low temperatures. It is usually described as the character of quantum spin fluctuations in strongly frustrated system.\cite{Ramirez, Nakatsuji, Okamoto, Terada9} Therefore, the specific-heat data at sub-Kelvin tempratures demonstrate that the ground state of CuFeO$_2$ is actually a co-existence of 4SL phase and spin fluctuated or short-range phase.

With Ga doping, the gapped excitation term disappears, indicating the weakness of the spin anisotropy. This is also consistent with the phase diagram's indication that upon doping the ground state is changed from the gapped 4SL to un-gapped FEIC. On the other hand, the power-law term is strongly enhanced with increasing $x$ to 0.08, and there are comparably strong magnetic excitations in the OPD phase of the $x =$ 0.08 and 0.12 samples. This indicates that the spin fluctuations of this material can be significantly enhanced when the low-$T$ phase is changed to OPD by doping with non-magnetic impurities.

\subsection{$\kappa(T)$ in zero field}

\begin{figure}
\includegraphics[clip,width=6.5cm]{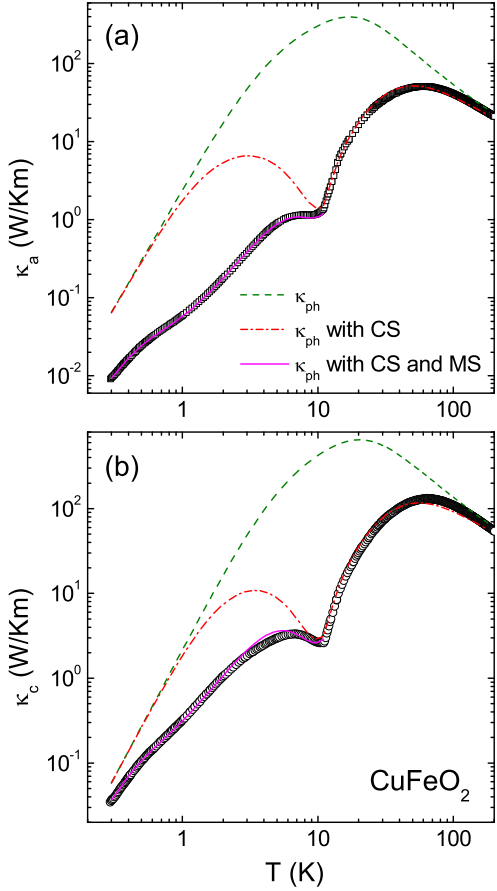}
\caption{(Color online) Temperature dependence of $\kappa_a$ (a) and $\kappa_c$ (b) of CuFeO$_2$ single crystals in zero field. The dashed lines show the calculated results using the Debye model with the magnetic scattering switched off, and displays a standard behavior of phonon heat transport. The dash-dot lines show the calculations using the Debye model with the scattering effect caused by the critical fluctuations of magnetic transition (CS). The solid lines are the fittings to low-temperature $\kappa$ using the Debye model including both the critical fluctuation scattering and the magnetic-excitation scattering (MS).}
\end{figure}

Figure 3 shows the $a$-axis and $c$-axis thermal conductivities of the $x$ = 0 samples in zero field. It is notable that the high-$T$ thermal conductivities actually exhibit rather large magnitudes compared with some other insulators. CuFeO$_2$ is known to be a good semiconductor,\cite{Rogers, Dordor} so it is necessary to estimate the electronic thermal conductivity $\kappa_e$ by using the Wiedemann-Franz law $\kappa_e = LT/ \rho$, where $\rho$ is the electrical resistivity and $L$ (= 2.44 $\times$ 10$^{-8}$ W$\Omega$/K$^2$) is the Lorenz number. Based on the resistivity data reported in the literature,\cite{Rogers} $\kappa_e$ along the $a$ and $c$ axes at 200 K can be estimated to be 3.7 $\times$ 10$^{-4}$ and 1.9 $\times$ 10$^{-8}$ W/Km, respectively. Apparently, the electronic contribution to $\kappa$ is negligibly small and the thermal conductivity is mainly the phononic term. With lowering temperature, however, the $\kappa(T)$ curves do not exhibit the large phonon peak, usually located at 10--20 K.\cite{Berman} Instead, they show broad peaks centering at $\sim$ 60 K, below which the temperature dependence of $\kappa$ are rather complicated. As shown in Fig. 3, both $\kappa_a(T)$ and $\kappa_c(T)$ show a slight change of slope at $T_{N1}$ (= 14 K) and a remarkable dip-like feature at $T_{N2}$ (= 11 K). Ultrasonic measurement has revealed previously that the phonon velocity shows a sharp minimum (with 5--6\% change) at $T_{N1}$ and a step increase (1--2\%) at $T_{N2}$, respectively.\cite{Quirion1, Quirion2} Apparently, the changes of velocity are not big enough to be responsible for the anomalies of $\kappa$. The dip of $\kappa(T)$ should be caused by a drastic phonon scattering by the critical magnetic fluctuations at the magnetic transition, which significantly changes the phonon mean free path.\cite{Wu_CHC, Zhao_CVO} In order to identify this kind of phonon scattering mechanism, we tried to fit the experimental data based on the classical Debye model. The phonon thermal conductivity can be expressed as\cite{Berman, Ziman}
\begin{equation}\label{eq:eps}
\kappa_{ph}=\frac{k_B}{2\pi^2 v_p}\left(\frac{k_B}{\hbar}\right)^3
T^3\int_0^{\Theta_D/T} \frac{x^4e^x}{(e^x-1)^2} \tau(\omega,T)dx,
\end{equation}
in which $x = \hbar\omega/k_BT$, $\omega$ is the phonon frequency, $\Theta_D$ is the Debye temperature, and $\tau(\omega,T)$ is the mean lifetime or scattering rate of phonons. The phonon relaxation is usually defined as
\begin{equation}\label{eq:eps}
\tau^{-1} = v_p/L + A\omega^4 + BT\omega^3\exp(-\Theta_D/bT) +   \tau_{m}^{-1},
\end{equation}
where the four terms represent the phonon scattering by the grain boundary, scattering by the point defects, the phonon-phonon Umklapp scattering, and the magnetic scattering associated with magnetic phase transitions, respectively. $\Theta_D$ is the Debye temperature, the phonon velocity $v_p$ is calculated from the equation $v_p = \Theta_D({k_B}/{\hbar})(6\pi^2 n)^{-1/3}$, where $n$ is the number density of atoms, $L$ is the sample width, and $A$, $B$ and $b$ are adjustable parameters. According to Kawasaki's phenomenological theory,\cite{Kawasaki, Rivers} the critical phonon scattering at the magnetic transition can be expressed as
\begin{equation}\label{eq:eps}
\tau_{c}^{-1} = C \omega^2 T\left[D(1-T/T_c)^{\alpha}+\omega\right]
\end{equation} for $T > T_c$, and
\begin{equation}\label{eq:eps}
\tau_{c}^{-1} = C \omega^2 T\left[D(T/T_c-1)^{\alpha'}+\omega\right]
\end{equation} for $T < T_c$.
Here, $C$, $D$, $\alpha$, and $\alpha'$ are free parameters. $T_c$ is the critical temperature corresponding to the dip position of the $\kappa(T)$ curve, which is selected as 11 K. Using these formulas, the $\kappa(T)$ data are fitted. The best fitted results, shown as the dash-dot lines in Fig. 3, were chosen to reproduce the high-$T$ behavior of $\kappa_a$ ($\kappa_c$) with parameters $L = $ 3.73 $\times$ 10$^{-43}$ (3.35 $\times$ 10$^{-43}$) m, $A =$ 1.6 $\times$ 10$^{-43}$ (7.5 $\times$ 10$^{-44}$) s$^3$, $B =$ 3.1 $\times$ 10$^{-31}$ (1.05 $\times$ 10$^{-31}$) s$^2$K$^{-1}$, $b =$ 2.8 (2.8), $C =$ 2.5 $\times$ 10$^{-4}$ (1.2 $\times$ 10$^{-4}$) s$^2$K, $D = $ 4.8 $\times$ 10$^{13}$ (4.7 $\times$ 10$^{13}$) s$^{-1}$, $\alpha =$ 1.4 (1.4), $\alpha' =$ 2.5 (2.6). However, using these formulas, the calculated $\kappa$ for $T < T_c$ are still much larger than the experimental data. Apparently, there should be another phonon-scattering term at temperatures below $T_c$. It is worth mentioning that both the $a$- and $c$-axis $\kappa(T)$ for $T < T_c$ display a change of slope at about 1 K. This anomaly has some correspondence to the $C_m(T)$ data. As discussed above, there seem to be two kinds of magnetic excitations in magnetic ordering phase, that is, the exponential magnetic excitations (magnons) co-exist with power-law magnetic excitations at sub-Kelvin temperatures. In this regard, there is no theoretical formula to describe the phonon scattering rate by these low-energy magnetic excitations. We propose a phenomenological expression of this additional magnetic scattering term at $T < T_c$ by taking into account the magnetic specific heat,
\begin{equation}\label{eq:eps}
\tau_{m'}^{-1} = m(T\omega)^{-1/2}C_{m1} + n(T\omega)^{-1/2}\exp(-T_c/dT)C_{m2}
\end{equation}
where $m$, $n$, $d$ are free parameters, and $C_{m1}$ and $C_{m2}$ are the power-law and exponential terms of the magnetic specific heat, respectively. The low-$T$ $\kappa_a$ ($\kappa_c$) data can be fitted quite well with parameters $m =$ 2.15 $\times$ 10$^{16}$ (2.1 $\times$ 10$^{15}$) J$^{-1}$K$^{3/2}$s$^{-3/2}$mol, $n =$ 5.72 $\times$ 10$^{14}$ (7.0 $\times$ 10$^{13}$) J$^{-1}$K$^{3/2}$s$^{-3/2}$mol, and $d =$ 1.31 (1.25). The parameters $m$ and $n$ in $\kappa_a$ are clearly larger than those in $\kappa_c$, which indicates a stronger phonon scattering by magnetic excitations for the in-plane phonon transport. Finally, if we switch off all the magnetic scattering by setting $\tau_{c}^{-1}$ and $\tau_{m'}^{-1}$ to zero, the calculated $\kappa(T)$ show a typical behavior of phonon transport with very large phonon peaks, as shown by the dashed lines in Fig. 3.

\begin{figure}
\includegraphics[clip,width=6.5cm]{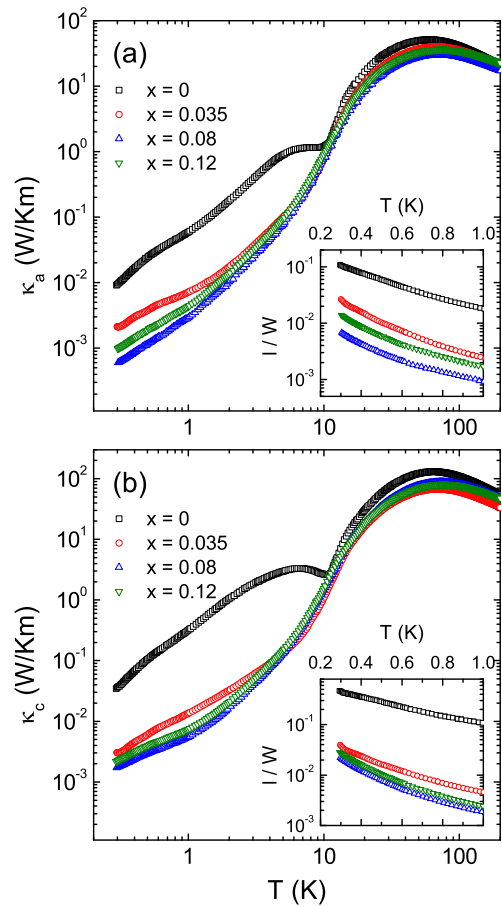}
\caption{(Color online) Temperature dependence of $\kappa_a$ (a) and $\kappa_c$ (b) of CuFe$_{1-x}$Ga$_x$O$_2$ ($x =$ 0, 0.035, 0.08, and 0.12) single crystals in zero field. The insets show the temperature dependence of the phonon mean free path $l$ divided by the averaged sample width $W$ at sub-Kelvin temperatures for all samples.}
\end{figure}

Figure 4 shows the $a$-axis and $c$-axis thermal conductivities of Ga-doped CuFe$_{1-x}$Ga$_x$O$_2$ ($x =$ 0.035, 0.08, and 0.12) in comparison with those of $x$ = 0 samples. Although the doping dependence of $\kappa_a$ and $\kappa_c$ are somewhat different at very low temperatures, the main phenomenon is that the low-$T$ thermal conductivities are strongly suppressed in the Ga-doped samples. It should be noted that the impurity doping deteriorates the periodicity of the crystal lattice and therefore shortens the mean free path of phonons, giving rise to the reduction of $\kappa$. This is the main reason that $\kappa$ at high temperatures (above several Kelvins) are weakened with Ga doping. However, this kind of phonon scattering caused by impurities is negligible at very low temperatures because the wave-length of the phonon is long enough to far exceed the size of local lattice distortion.\cite{Berman} Considering the specific-heat data upon Ga doping, the clear suppression of low-$T$ $\kappa$ in Ga-doped samples should be caused by a stronger phonon scattering by the magnetic fluctuation. The phonon thermal conductivity can be expressed as a kinetic formula $\kappa = \frac{1}{3}C_{ph} v l$, in which $C_{ph} = \beta T^3$ is the low-$T$ phononic specific heat, $v$ is the averaged sound velocity and is nearly $T$-independent at low temperatures, and $l$ is the mean free path of phonons.\cite{Berman} With decreasing temperature, the microscopic scatterings of phonons are gradually smeared out and the $l$ increases continuously until it reaches the averaged sample width $W = 2\sqrt{A/\pi}$, where $A$ is the cross-section area of sample.\cite{Berman} This boundary scattering limit of phonons can be achieved only at very low temperatures and the $T$-dependence of $\kappa$ is the same as the $T^3$ law of the specific heat \cite{Berman, Ziman}. In the present case, none of the low-$T$ $\kappa(T)$ curves shows the $T^3$ dependence at sub-Kelvin temperatures, which means that the microscopic scattering is not negligible even at temperatures as low as 0.3 K. With the $\Theta_D$ value (= 180 K) from the specific-heat data, the phonon mean free path can be calculated assuming that $\kappa$ is purely phononic.\cite{Zhao_NCO, Sun_Comment} The insets to Fig. 4 show the temperature dependence of the ratio $l/W$. It is found that for $x =$ 0 sample, the $l/W$ ratios at the lowest temperature are only about 0.1 and 0.5 for $\kappa_a$ and $\kappa_c$, respectively. This indicates that there must be a magnetic scattering effect at low temperatures, which is more significant for phonons transporting along the $a$ axis. In Ga-doped samples, it is clear that the magnetic scattering effect is so strongly enhanced that the phonon mean free paths are much smaller than those of the undoped samples. This is compatible with the low-$T$ specific heat data that demonstrate much stronger magnetic fluctuations in Ga-doped samples.

It is also notable that $\kappa$ of undoped samples are rather anisotropic, that is, $\kappa_a$ are several times smaller than $\kappa_c$ at low temperatures. It seems that the magnetic excitations are anisotropic and scatter the in-plane phonons more effectively. With Ga doping, the anisotropy of $\kappa$ becomes much weaker, which means that the magnetic anisotropy is weakened by the nonmagnetic impurities.

\subsection{$\kappa(H)$ and $\kappa(T)$ of CuFeO$_2$}

\begin{figure}
\includegraphics[clip,width=8.5cm]{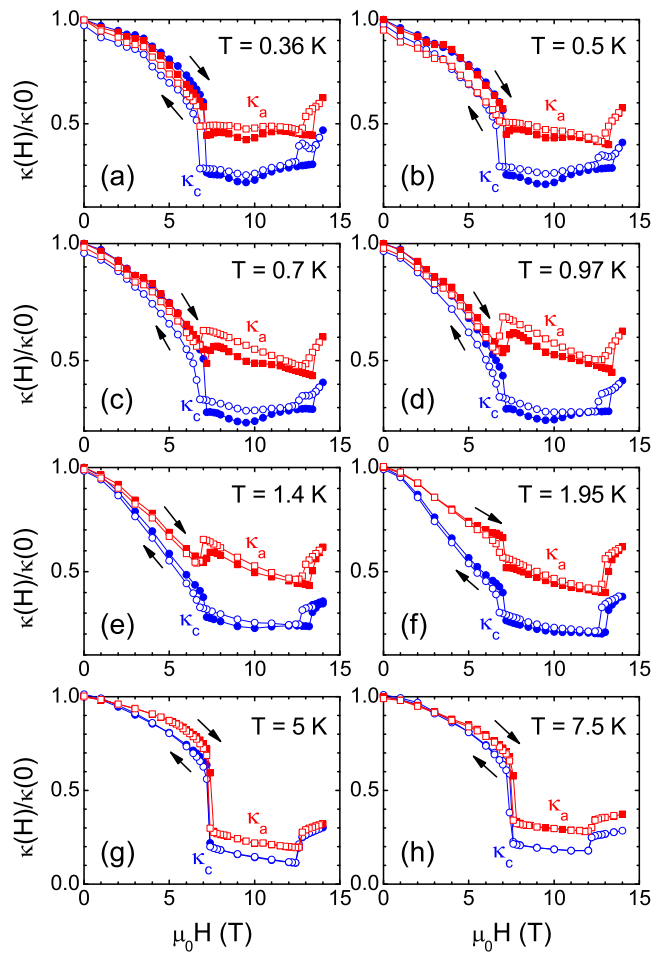}
\caption{(Color online) Magnetic-field dependence of the $\kappa_a$ and $\kappa_c$ of CuFeO$_2$ single crystals in $H \parallel c$ after ZFC. As indicated by the arrows, the data shown with solid symbols are measured in the field-up process, while the open symbols show the data in the field-down process.}
\end{figure}

Figure 5 shows the magnetic-field dependence of $\kappa_a$ and $\kappa_c$ at low temperatures with zero-field cooling (ZFC). The magnetic field was applied along the $c$ axis. Both $\kappa_a(H)$ and $\kappa_c(H)$ isotherms display complex field dependence: $\kappa$ gradually decreases with increasing field with a sudden change at $\mu_0H_{c1} \sim$ 7 T, then becomes weakly field dependent until another drastic change at $\mu_0H_{c2} \sim$ 13 T, followed by an increase with further increasing field. The two characteristic fields are corresponding to the magnetic transitions from the 4SL to FEIC phase and from the FEIC to 5SL phase, respectively.\cite{Kimura2} Below $H_{c1}$, the Fe$^{3+}$ spins are mainly ordered in the gapped 4SL phase,\cite{Fishman, Haraldsen} and the decrease of $\kappa$ is probably caused by the enhancement of the magnon scattering of phonons with increasing field. This explanation is reasonable since the anisotropy gap decreases linearly with increasing field and magnetic excitations are gradually populated. Between $H_{c1}$ and $H_{c2}$, the Fe$^{3+}$ spins are mainly ordered in the gapless FEIC phase,\cite{Fishman, Haraldsen} and there are large amount of magnons that can strongly scatter phonons, so $\kappa$ is strongly suppressed and has weak field dependence. Above $H_{c2}$, the Fe$^{3+}$ spins form the gapped 5SL phase, which has weak magnon excitations because of the re-opening of a gap.\cite{Fishman, Haraldsen} As a result, the phonon scattering is weakened and $\kappa$ recovers at $H > H_{c2}$.

A notable feature of the $\kappa(H)$ isotherms is the difference in the sudden changes of $\kappa$ at $H_{c1}$. At $T \ge$ 1.95 K and $T \le$ 0.5 K, both $\kappa_a$ and $\kappa_c$ show a similar step-like decrease, which corresponds well to the jump changes of magnetostriction at critical fields.\cite{Kimura2} In the intermediate region of 0.7 K $\le T \le$ 1.4 K, however, $\kappa_c$ still shows a step-like decrease while $\kappa_a$ shows a step-like increase at $H_{c1}$. Apparently, the main difference between $\kappa_a(H)$ and $\kappa_c(H)$ is likely due to the competing roles of magnons in heat transport, that is, acting as heat carries or phonon scatterers. Because of the strong anisotropy,\cite{Terada7, Terada8, Ye1} the magnons are more dispersive along the $ab$ plane. Therefore, the strong suppression of $\kappa_c(H)$ at $H_{c1}$ is due to the phonon scattering by the weak dispersive magnons in the $c$ direction, while for $\kappa_a(H)$, the in-plane magnons can not only scatter phonons but also transport heat.

Another phenomenon is that the $\kappa(H)$ isotherms show clear irreversibility in the whole field range at $T \le$ 1.95 K and the hysteresis becomes more pronounced with decreasing temperature. Note that the magnetization of CuFeO$_2$ also exhibits irreversibility in the whole field range.\cite{Kimura2} The difference between the $M(H)$ and $\kappa(H)$ is that the former can be irreversible at rather high temperatures while the latter are irreversible only at $T <$ 2 K. It is intriguing that $\kappa$ with decreasing field is smaller than with increasing field at $H > H_{c1}$, but the relative magnitudes become reverse at $H < H_{c1}$. In the measurement of magnetostriction with field up to 14 T, hysteretic behavior has been observed,\cite{Kimura2} which is consistent with the $\kappa(H)$ data. The change of magnetostriction could affect the phonon spectrum and may lead to the irreversible behavior of $\kappa(H)$. However, previous reserch found that the magnetostriction either parallel to or perpendicular to the $c$ axis shows monotonic changes with increasing field,\cite{Kimura2} which is very different from the changes of $\kappa$.

\begin{figure}
\includegraphics[clip,width=6cm]{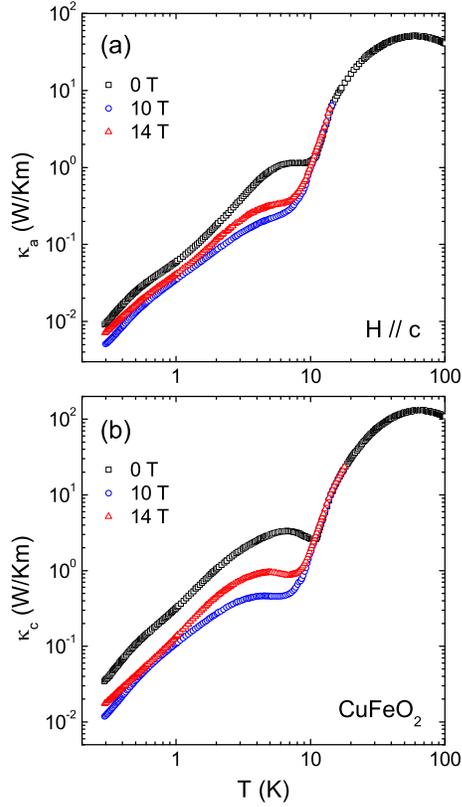}
\caption{(Color online) Temperature dependence of the $\kappa_a$ and $\kappa_c$ of CuFeO$_2$ single crystals in $H \parallel c$. Note that 0, 10 and 14 T correspond to the 4SL, FEIC and 5SL phases, respectively.}
\end{figure}

Figure 6 shows the temperature dependence of $\kappa_a$ and $\kappa_c$ with $H \parallel c$. Applying 10 T and 14 T fields along the $c$ axis, the dip feature at $T_{N2}$ becomes wider and moves to lower temperatures, which is consistent with the $T_{N2}$ vs $H$ shown in Fig. 1(a). At the same time, $\kappa$ at $T < T_{N2}$ are significantly suppressed in 10 T field but recover somewhat in 14 T field, as also indicated by the $\kappa(H)$ data. The 1 K anomaly in the zero-field $\kappa(T)$ curves is almost unchanged in 10 and 14 T fields. We discussed above that the 1 K anomaly is likely related to the power-law magnetic excitations at very low temperatures. The consistency of the anomaly indicates that these kinds of magnetic excitations are not significantly changed when a magnetic field drives the phase transition from the 4SL to the FEIC and then to the 5SL.

\subsection{$\kappa(H)$ and $\kappa(T)$ of the $x =$ 0.035 samples}

\begin{figure}
\includegraphics[clip,width=8.5cm]{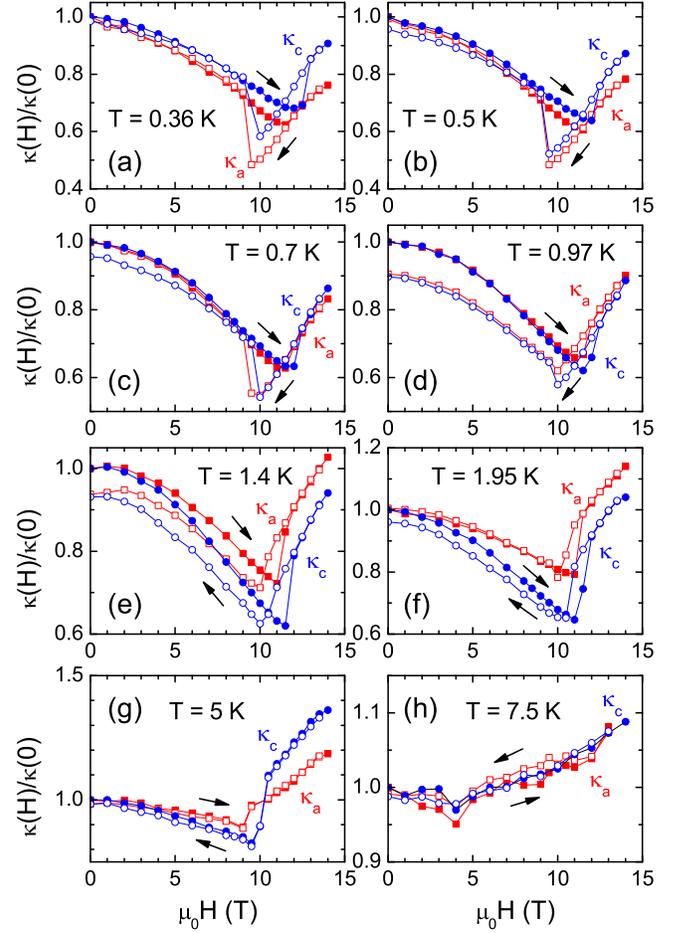}
\caption{(Color online) Magnetic-field dependence of $\kappa_a$ and $\kappa_c$ of the $x =$ 0.035 single crystals in $H \parallel c$ after ZFC. As indicated by the arrows, the data shown with solid symbols are measured in the field-up process, while the open symbols show the data in the field-down process.}
\end{figure}

Figure 7 shows the magnetic-field dependence of $\kappa_a$ and $\kappa_c$ at low temperatures for the $x =$ 0.035 single crystals with $H \parallel c$, which are significantly different from that of the undoped CuFeO$_2$.  First, the changes of $\kappa$ with field become weaker than those of the undoped samples. Second, the field dependence is rather similar between the $\kappa_a$ and $\kappa_c$, indicating that the Ga doping weakens the magnetic anisotropy. This similarity is consistent with what the $\kappa(T)$ data indicate. Third, the field dependence of $\kappa$ is qualitatively different from those of the undoped samples, due to the differences in the ground states and field-induced transitions.

The main feature of the $\kappa(H)$ isotherms at $T \le$ 5 K is a minimum located at the magnetic transition from the FEIC to 5SL phase (the transition field is defined as $H_{c2}$). In addition, there is also a large irreversibility of the $\kappa(H)$ curves in the $x =$ 0.035 samples. This irreversibility mainly appears near the magnetic transition for both $\kappa_a$ and $\kappa_c$, which results in obviously different minimum-fields for field sweeping up and down. It is notable that most of the $\kappa(H)$ curves show a step-like increase at $H_{c2}$ for an increasing field, but show a step-like decrease at $H_{c2}$ for decreasing field, particularly at very low temperatures. This behavior is qualitatively different from that of the $x =$ 0 samples at the transition from the FEIC to 5SL phase (see Fig. 4). As a result, near the transition from the FEIC to 5SL phase, the $\kappa(H)$ curves of the $x =$ 0 samples show a rectangular single-loop while those of the $x =$ 0.035 samples show a butterfly-shaped loop.

\begin{figure}
\includegraphics[clip,width=6cm]{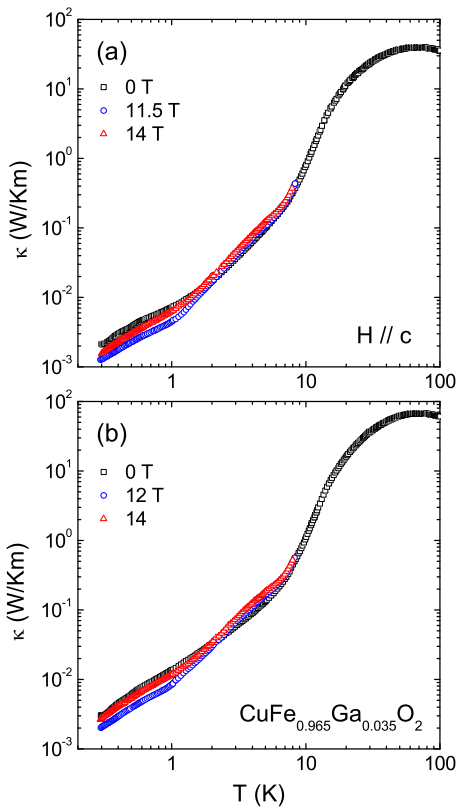}
\caption{(Color online) Temperature dependence of the $\kappa_a$ and $\kappa_c$ of the $x =$ 0.035 single crystals in $H \parallel c$. The transition field between the FEIC ground state and the high-field 5SL phase is about 11.5 T.}
\end{figure}

Figure 8 shows the temperature dependence of $\kappa_a$ and $\kappa_c$ with $H \parallel c$. Applying 11.5 (or 12) T and 14 T fields along the $c$ axis, the shoulder feature at $T_{N2}$ does not change much, which is consistent with the phase diagram shown in Fig. 1(b). At 11.5 (or 12) T, near the critical field, the change of the $\kappa(T)$ slope becomes much sharper and the curves display a kink-like feature at about 1 K, which corresponds to the boundary between the FEIC and 5SL phases. The main reason for this enhanced anomaly should be the phonon scattering by the critical fluctuations.

As already mentioned above, the low-$T$ $\kappa$ of the $x =$ 0.035 samples are significantly smaller than those of the undoped samples. Both the weak temperature dependence (compared to the $T^3$ law of the boundary scattering limit) and the short mean free path of phonons indicate that the phonons are strongly scattered even at sub-Kelvin temperatures, which can only be attributed to the magnetic scattering effect. This is consistent with the specific heat data, showing that the magnetic excitations or fluctuations at low temperatures are strongly enhanced with Ga doping.

\subsection{$\kappa(H)$ and $\kappa(T)$ of the $x =$ 0.08 samples}

\begin{figure}
\includegraphics[clip,width=6cm]{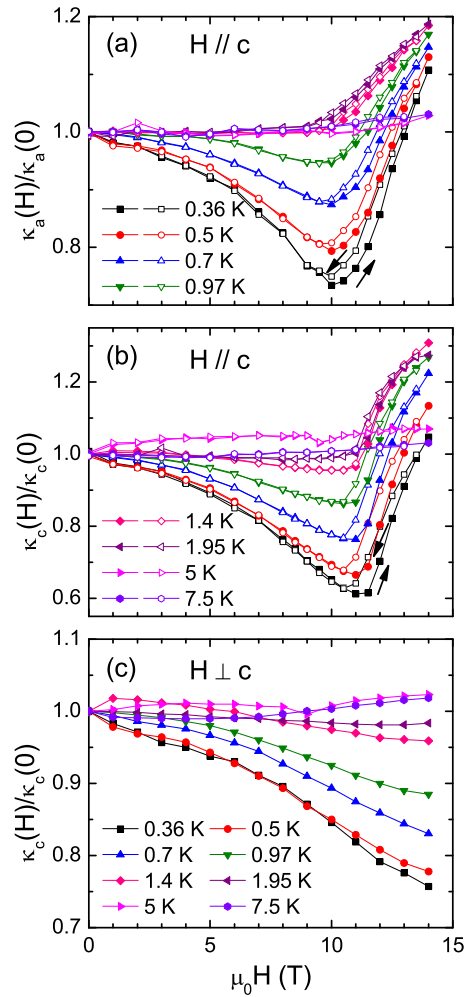}
\caption{(Color online) (a,b) Magnetic-field dependence of $\kappa_a$ and $\kappa_c$ of the $x =$ 0.08 single crystals with $H \parallel c$ and at low temperatures after ZFC. As indicated by the arrows, the data shown with solid symbols are measured in the field-up process, while the open symbols show the data in the field-down process. (c) Magnetic-field dependence of $\kappa_c$ with $H \perp c$. In this case the data are reversible.}
\end{figure}

Figures 9(a) and 9(b) show the magnetic-field dependence of $\kappa_a$ and $\kappa_c$ at low temperatures and in the $c$-axis field for the $x =$ 0.08 samples. At $T \le$ 1.95 K, the $\kappa_a(H)$ and $\kappa_c(H)$ isotherms show a valley-like feature: $\kappa$ gradually decreases with increasing field and arrives a minimum at $H_c \sim$ 11 T, followed by a quick increase at $H > H_c$. In addition, an irreversibility is observed above $H_c$ between the field-up and field-down curves. Note that these behaviors are somewhat different from the $\kappa(H)$ of the $x =$ 0.035 samples in two aspects. First, the $\kappa(H)$ of the $x =$ 0.08 samples do not show sharp decrease or increase near the minimum values. Second, the irreversibility exists over all the high-field region and does not appear at the low-field side of the minimum. This is understandable since the ground states are different between the $x =$ 0.035 and 0.08 samples.

These behaviors indicate a kind of spin-structure transition driven by the $c$-axis field near 10 T. The 8\%-Ga doped CuFeO$_2$ is known to have the OPD ground state but the possible field-induced transitions have not been studied in earlier works. Analogous to the case of some other low-dimensional magnets, such as the zigzag-chain material CoNb$_2$O$_6$,\cite{Nakajima4} it is likely that magnetic field drives the OPD phase to some kind of ferrimagnetic phase.

For comparison, the $\kappa_c$ isotherms with $H \perp c$ are shown in Fig. 9(c). At low temperatures, they exhibit a simple decrease of $\kappa$ with increasing field without any transition and irreversibility in field up to 14 T. However, it should be noted that the $\kappa(H)$ behavior with $H \perp c$ is very similar to the low-field behavior with $H \parallel c$, which may indicate a similar field-induced transition at a higher field.

\begin{figure}
\includegraphics[clip,width=6cm]{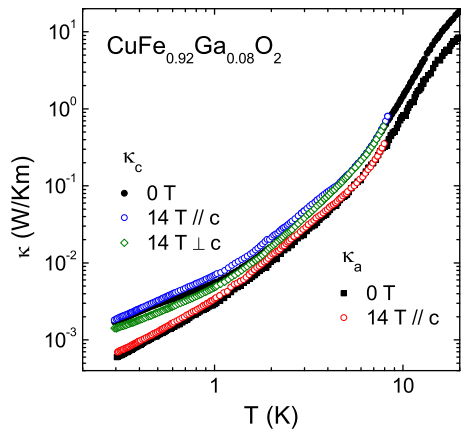}
\caption{(Color online) Temperature dependence of $\kappa_a$ and $\kappa_c$ of the $x =$ 0.08 single crystals in zero field and 14 T $\parallel c$ or $\perp c$.}
\end{figure}

Figure 10 shows the temperature dependence of $\kappa_a$ and $\kappa_c$ with 14 T along or perpendicular to the $c$ axis. With 8\%-Ga doping, the 1 K anomaly in the zero-field $\kappa(T)$ curves also moves to a bit lower temperature, similar to the case of the $x =$ 0.035 samples. The 14 T fields in all directions have weak impact on this feature.

\subsection{$\kappa(H)$ and $\kappa(T)$ of the $x =$ 0.12 samples}

\begin{figure}
\includegraphics[clip,width=6cm]{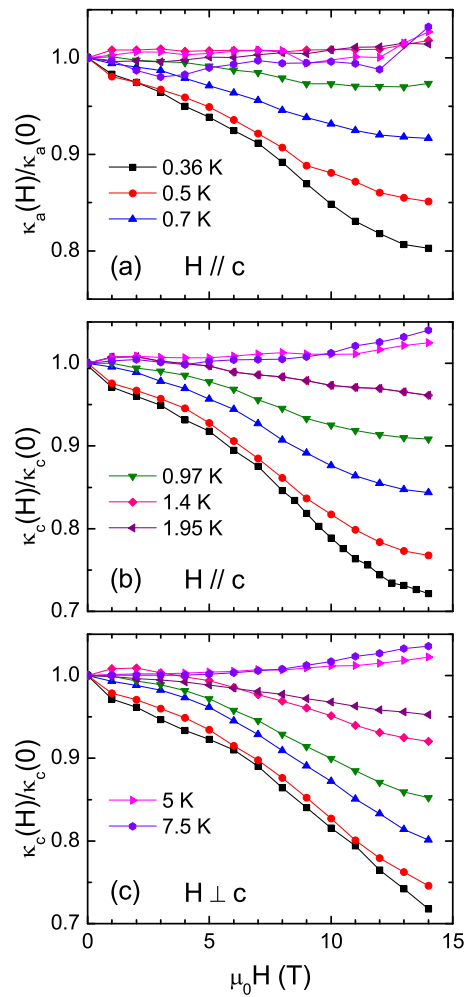}
\caption{(Color online) (a,b) Magnetic-field dependence of the $\kappa_a$ and $\kappa_c$ of the $x =$ 0.12 single crystals with $H \parallel c$ and at low temperatures. (c) Magnetic-field dependence of $\kappa_c$ with $H \perp c$.}
\end{figure}

Figure 11 shows the magnetic-field dependence of $\kappa_a$ and $\kappa_c$ at low temperatures for the $x =$ 0.12 samples. All the $\kappa(H)$ isotherms show similar behavior, that is, $\kappa$ gradually decreases with increasing field up to 14 T. Furthermore, these $\kappa(H)$ curves are similar to those of the $x =$ 0.08 sample with $H \perp c$. The ground state and phase diagram of the 12\%-Ga doped CuFeO$_2$ have not been previously reported. Based on the phase diagram with $x$ up to 0.08, as shown in Fig. 1(c), it is likely that the ground state of the $x =$ 0.12 sample is also OPD. This is supported by the nearly identical specific heat data between the $x =$ 0.08 and 0.12 samples, as shown in Fig. 2(b). However, it seems that if the magnetic phase transitions of the $x =$ 0.12 samples exists, they might happen at very high field for either $H \parallel c$ or $H \perp c$.

\begin{figure}
\includegraphics[clip,width=6cm]{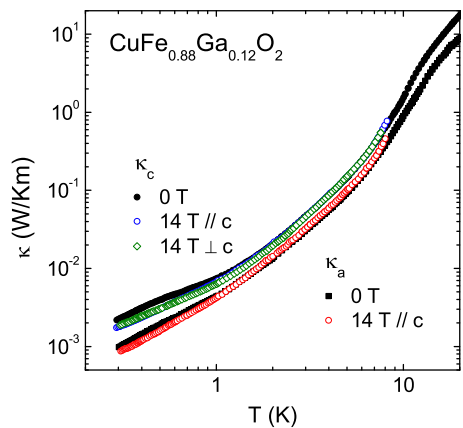}
\caption{(Color online) Temperature dependence of the $\kappa_a$ and $\kappa_c$ of the $x =$ 0.12 single crystals in zero field and 14 T $\parallel c$ or $\perp c$.}
\end{figure}

Figure 12 shows the temperature dependence of $\kappa_a$ and $\kappa_c$ with 14 T along or perpendicular to the $c$ axis. With 12\%-Ga doping, the 1 K anomaly is still observable in the zero-field $\kappa(T)$ curves. The 14 T fields in all directions weaken this feature.

\section{SUMMARY}

In this work, we study the thermal conductivity of CuFe$_{1-x}$Ga$_x$O$_2$ ($x =$ 0--0.12) single crystals at temperatures down to 0.3 K and in magnetic fields up to 14 T. It was found that the thermal conductivities show drastic anomalies at temperature- or field-induced magnetic transitions, pointing to a strong spin-phonon coupling in this material. The temperature dependence of $\kappa$ is rather complicated at very low temperatures and indicates magnetic scattering of phonons, which reveals non-negligible magnetic fluctuations in the ``ground state" of pure and Ga-doped samples. This phenomenon is also evidenced by the specific-heat data at temperatures down to 0.4 K. In addition, the low-$T$ $\kappa(H)$ isotherms exhibit irreversibility in a broad region of magnetic field, which is not completely understood and calls for further investigation.

\begin{acknowledgements}

This work was supported by the National Natural Science Foundation of China (Grant Nos. 11374277, 11574286, U1532147, 11404316), the National Basic Research Program of China (Grant Nos. 2015CB921201, 2016YFA0300103), and the Opening Project of Wuhan National High Magnetic Field Center (Grant No. 2015KF21).

\end{acknowledgements}


\begin{thebibliography}{}

\bibitem{Schmid}
H. Schmid, Ferroelectrics {\bf 162}, 317 (1994).

\bibitem{Kimura1}
T. Kimura, T. Goto, H. Shintani, K. Ishizaka, T. Arima, and Y. Tokura, Nature {\bf 426}, 55 (2003).

\bibitem{Hur}
N. Hur, S. Park, P. A. Sharma, J. S. Ahn, S. Guha, and S-W. Cheong, Nature {\bf 429}, 392 (2004).

\bibitem{Lottermoser}
T. Lottermoser, T. Lonkai, U. Amann, D. Hohlwein, J. Ihringer, and M. Fiebig, Nature {\bf 430}, 541 (2004).

\bibitem{MF_Review}
S.-W. Cheong and M. Mostovoy, Nat. Mater. {\bf 6}, 13 (2007).

\bibitem{Kimura2}
T. Kimura, J. C. Lashley, and A. P. Ramirez, Phys. Rev. B {\bf 73}, 220401(R) (2006).

\bibitem{Arima}
T. Arima, J. Phys. Soc. Jpn. {\bf 76}, 073702 (2007).

\bibitem{Muir}
A. H. Muir Jr. and H. Wiedersich, J. Phys. Chem. Solids, {\bf 28}, 65 (1967).

\bibitem{Kawamura}
H. Kawamura, J. Phys.: Condens. Matt. {\bf 10}, 4707 (1998).

\bibitem{Mitsuda1}
S. Mitsuda, H. Yoshizawa, N. Yaguchi, and M. Mekata, J. Phys. Soc. Jpn. {\bf 60}, 1885 (1991).

\bibitem{Mitsuda2}
S. Mitsuda, N. Kasahara, T. Uno, and M. Mase, J. Phys. Soc. Jpn. {\bf 67}, 4026 (1998).

\bibitem{Ye1}
F. Ye, J. A. Fernandez-Baca, R. S. Fishman, Y. Ren, H. J. Kang, Y. Qiu, and T. Kimura, Phys. Rev. Lett. {\bf 99}, 157201 (2007).

\bibitem{Terada1}
N. Terada, S. Mitsuda, H. Ohsumi, and K. Tajima, J. Phys. Soc. Jpn. {\bf 75}, 023602 (2006).

\bibitem{Ye2}
F. Ye, Y. Ren, Q. Huang, J. A. Fernandez-Baca, P. C. Dai, J. W. Lynn, and T. Kimura, Phys. Rev. B {\bf 73}, 220404(R) (2006).

\bibitem{Terada2}
N. Terada, Y. Tanaka, Y. Tabata, K. Katsumata, A. Kikkawa, and S. Mitsuda, J. Phys. Soc. Jpn. {\bf 75}, 113702 (2006).

\bibitem{Petrenko}
O. A. Petrenko, G. Balakrishnan, M. R. Lees, D. McK. Paul, and A. Hoser, Phys. Rev. B {\bf 62}, 8983 (2000).

\bibitem{Terada3}
N. Terada, Y. Narumi, K. Katsumata, T. Yamamoto, U. Staub, K. Kindo, M. Hagiwara, Y. Tanaka, A. Kikkawa, H. Toyokawa, T. Fukui, R. Kanmuri, T. Ishikawa, and H. Kitamura, Phys. Rev. B {\bf 74}, 180404(R) (2006).

\bibitem{Lummen}
T. T. A. Lummen, C. Strohm, H. Rakoto, A. A. Nugroho, and P. H. M. van Loosdrecht, Phys. Rev. B {\bf 80}, 012406 (2009).

\bibitem{Seki1}
S. Seki, Y. Yamasaki, Y. Shiomi, S. Iguchi, Y. Onose, and Y. Tokura, Phys. Rev. B {\bf 75}, 100403(R) (2007).

\bibitem{Terada4}
N. Terada, T. Nakajima, S. Mitsuda, H. Kitazawa, K. Kaneko, and N. Metoki, Phys. Rev. B {\bf 78}, 014101 (2008).

\bibitem{Terada5}
N. Terada, T. Nakajima, S. Mitsuda, H. Kitazawa, J. Phys.: Conf. Ser. {\bf 145}, 012071 (2009).

\bibitem{Seki2}
S. Seki, H. Murakawa, Y. Onose, and Y. Tokura, Phys. Rev. Lett. {\bf 103}, 237601 (2009).

\bibitem{Terada6}
N. Terada, T. Kawasaki, S. Mitsuda, H. Kimura, and Y. Noda, J. Phys. Soc. Jpn. {\bf 74}, 1561 (2005).

\bibitem{Nakajima1}
T. Nakajima, S. Mitsuda, K. Kitagawa, N. Terada, T. Komiya, and Y. Noda, J. Phys.: Condens. Matt. {\bf 19}, 145216 (2007).

\bibitem{Hess}
C. Hess, Eur. Phys. J. Spec. Top. {\bf 151}, 73 (2007).

\bibitem{Sologubenko}
A. V. Sologubenko, T. Lorenz, H. R. Ott, and A. Friemuth, J. Low. Temp. Phys. {\bf 147}, 387 (2007).

\bibitem{Zhao_SG}
X. Zhao, Z. Y. Zhao, X. G. Liu, and X. F. Sun, Sci. China-Phys. Mech. Astron. {\bf 59}, 117501 (2016).

\bibitem{Yamashita}
M. Yamashita, N. Nakata, Y. Senshu, M. Nagata, H. M. Yamamoto, R. Kato, T. Shibauchi, and Y. Matsuda, Science {\bf 328}, 1246 (2010).

\bibitem{Sun_DTN}
X. F. Sun, W. Tao, X. M. Wang, and C. Fan, Phys. Rev. Lett. {\bf 102}, 167202 (2009).

\bibitem{Zhao_IPA}
Z. Y. Zhao, B. Tong, X. Zhao, L. M. Chen, J. Shi, F. B. Zhang, J. D. Song, S. J. Li, J. C. Wu, H. S. Xu, X. G. Liu, and X. F. Sun, Phys. Rev. B {\bf 91}, 134420 (2015).

\bibitem{Jeon}
Byung-Gu Jeon, B. Koteswararao, C. B. Park, G. J. Shu, S. C. Riggs, E. G. Moon, S. B. Chung, F. C. Chou, and K. H. Kim, Sci. Rep. {\bf 6}, 36970 (2016).

\bibitem{Berman}
R. Berman, {\it Thermal Conduction in Solids} (Oxford University Press, Oxford, 1976).

\bibitem{Ziman}
J. M. Ziman, {\it Electrons and Phonons: The Theory of Transport Phenomena in Solids} (Oxford University Press, 1960).

\bibitem{Ashcroft}
N. W. Ashcroft and N. D. Mermin, {\it Solid State Physics} (Harcourt Brace  College Publishers, 1976).

\bibitem{Wang_HMO}
X. M. Wang, C. Fan, Z. Y. Zhao, W. Tao, X. G. Liu, W. P. Ke, X. Zhao, and X. F. Sun. Phys. Rev. B {\bf 82}, 094405 (2010).

\bibitem{Wang_TMO}
X. M. Wang, Z. Y. Zhao, C. Fan, X. G. Liu, Q. J. Li, F. B. Zhang, L. M. Chen, X. Zhao, and X. F. Sun, Phys. Rev. B {\bf 86}, 174413 (2012).

\bibitem{Zhao_GFO}
Z. Y. Zhao, X. M. Wang, C. Fan, W. Tao, X. G. Liu, W. P. Ke, F. B. Zhang, X. Zhao, and X. F. Sun, Phys. Rev. B {\bf 83}, 014414 (2011).

\bibitem{Zhao_DFO}
Z. Y. Zhao, X. Zhao, H. D. Zhou, F. B. Zhang, Q. J. Li, C. Fan, X. F. Sun, and X. G. Li, Phys. Rev. B {\bf 89}, 224405 (2014).

\bibitem{Song_Growth}
J. D. Song, J. C. Wu, X. Rao, S. J. Li, Z. Y. Zhao, X. G. Liu, X. Zhao, and X. F. Sun, J. Cryst. Growth {\bf 446}, 79 (2016).

\bibitem{Tari}
A. Tari, {\it Specific Heat of Matter at Low Temperatures} (Imperial College Press, 2003).

\bibitem{Svoboda}
P. Svoboda, P. Javorsk\'{y}, M. Divi\v{s}, V. Sechovsk\'{y}, F. Honda, G. Oomi, and A. A. Menovsky, Phys. Rev. B {\bf 63}, 212408 (2001).

\bibitem{Hemberger}
J. Hemberger, M. Hoinkis, M. Klemm, M. Sing, R. Claessen, S. Horn, and A. Loidl, Phys. Rev. B {\bf 72}, 012420 (2005).

\bibitem{Janiceka}
P. Jan\'{i}\v{c}ek, \v{C}. Dra\v{s}ar, P. Lo\v{s}t¡¯\'{a}k, J. Vejpravov\'{a}, and V. Sechovsk\'{y}, Physica B {\bf 403}, 3553 (2008).

\bibitem{Quirion1}
G. Quirion, M. J. Tagore, M. L. Plumer, and O. A. Petrenko, Phys. Rev. B {\bf 77}, 094111 (2008).

\bibitem{Terada7}
N. Terada, S. Mitsuda, Y. Oohara, H. Yoshizawa, and H. Takei, J. Magn. Magn. Mater. {\bf 272-276}, e997 (2004).

\bibitem{Terada8}
N. Terada, S. Mitsuda, T. Fujii, and D. Petitgrand, J. Phys.: Condens. Matt. {\bf 19}, 145241 (2007).

\bibitem{Nakajima2}
T. Nakajima, S. Mitsuda, T. Haku, K. Shibata, K. Yoshitomi, Y. Noda, N. Aso, Y. Uwatoko, and N. Terada, J. Phys. Soc. Jpn. {\bf 80}, 014714 (2011).

\bibitem{Nakajima3}
T. Nakajima, A. Suno, S. Mitsuda, N. Terada, S. Kimura, K. Kaneko, and H. Yamauchi, Phys. Rev. B {\bf 84}, 184401 (2011).

\bibitem{Ramirez}
A. P. Ramirez, B. Hessen, and M. Winklemann, Phys. Rev. Lett. {\bf 84}, 2957 (2000).

\bibitem{Nakatsuji}
S. Nakatsuji, Y. Nambu, H. Tonomura, O. Sakai, S. Jonas, C. Broholm, H. Tsunetsugu, Y. M. Qiu, and Y. Maeno, Science {\bf 309}, 1697 (2005).

\bibitem{Okamoto}
Y. Okamoto, M. Nohara, H. Aruga-Katori, and H. Takagi, Phys. Rev. Lett. {\bf 99}, 137207 (2007).

\bibitem{Terada9}
N. Terada, Y. Tsuchiya, H. Kitazawa, T. Osakabe, N. Metoki, N. Igawa, and K. Ohoyama, Phys. Rev. B {\bf 84}, 064432 (2011).

\bibitem{Rogers}
D. B. Rogers, R. D. Shannon, C. T. Prewitt, and J. L. Gillson, Inorg. Chem. {\bf 10}, 723 (1971).

\bibitem{Dordor}
P. Dordor, J. P. Chaminade, A. Wichainchai, E. Marquestaut, J. P. Doumerc, M. Pouchard, and P. Hagenmuller, J. Solid State Chem. {\bf 75}, 105 (1988).

\bibitem{Quirion2}
G. Quirion, M. L. Plumer, O. A. Petrenko, G. Balakrishnan, and C. Proust, Phys. Rev. B {\bf 80}, 064420 (2009).

\bibitem{Wu_CHC}
J. C. Wu, J. D. Song, Z. Y. Zhao, J. Shi, H. S. Xu, J. Y. Zhao, X. G. Liu, X. Zhao, and X. F. Sun, J. Phys.: Condens. Matt. {\bf }28, 056002 (2016).

\bibitem{Zhao_CVO}
X. Zhao, J. C. Wu, Z. Y. Zhao, Z. Z. He, J. D. Song, J. Y. Zhao, X. G. Liu, X. F. Sun, and X. G. Li, Appl. Phys. Lett. {\bf 108}, 242405 (2016).

\bibitem{Kawasaki}
K. Kawasaki, Prog. Theor. Phys. {\bf 29}, 801 (1963).

\bibitem {Rivers}
J. E. Rives, D. Walton, and G. S. Dixon, J. Appl. Phys. {\bf 41}, 1435 (1971).

\bibitem{Zhao_NCO}
Z. Y. Zhao, X. M. Wang, B. Ni, Q. J. Li, C. Fan, W. P. Ke, W. Tao, L. M. Chen, X. Zhao, and X. F. Sun, Phys. Rev. B {\bf 83}, 174518 (2011).

\bibitem{Sun_Comment}
X. F. Sun and Y. Ando, Phys. Rev. B {\bf 79}, 176501 (2009).

\bibitem{Fishman}
R. S. Fishman, F. Ye, J. A. Fernandez-Baca, J. T. Haraldsen, and T. Kimura, Phys. Rev. B {\bf 78}, 140407(R) (2008).

\bibitem{Haraldsen}
J. T. Haraldsen, R. S. Fishman, and G. Brown, Phys. Rev. B {\bf 86}, 024412 (2012).

\bibitem{Nakajima4}
T. Nakajima, S. Mitsuda, H. Okano, Y. Inomoto, and S. Kobayashi, K. Prokes, S. Gerischer, and P. Smeibidl, J. Phys. Soc. Jpn. {\bf 83}, 094723 (2014).


\end{thebibliography}
\end{document}